# Allocation des ressources radio en LTE

BENDAOUD Fayssal
Laboratoire de Télécommunication Tlemcen (LTT)
Université Abou Bekr Belkaid, Tlemcen, Algerie

ABDENNEBI Marwen
Laboratoire de Traitement et Transport de l'Information (L2TI)
Université de Paris13, Paris, France

DIDI fedoua
Laboratoire de Télécommunication Tlemcen (LTT)
Université Abou Bekr Belkaid, Tlemcen, Algerie

*Résumé*—

**Dans cet article nous nous intéressons à une tâche importante de l'eNodeB dans l'architecture des réseaux LTE, c'est le RRM (Radio Resource Management) son objectif est d'accepter ou de rejeter les demandes de connexion au réseau, en assurant une distribution optimale des ressources radio entre les UEs (Users Equipements). Il est constitué principalement de deux éléments AC (Admission Control) et PS (Packet Scheduling). Dans ce travail on se focalisera sur le PS, qui réalise une allocation efficace des ressources radio dans les deux sens c'est-à-dire Uplink (considéré dans notre cas) et Downlink.**

**Plusieurs approches et algorithmes ont été proposés dans la littérature pour répondre à ce besoin (allouer les ressources efficacement), cette diversité et multitude d'algorithmes est liée aux facteurs considérés permettant la gestion optimale de ressource radio, spécifiquement le type de trafic et QoS demandée par l'UE.**

**Dans cet article, une étude de plusieurs algorithmes d'ordonnancement proposé pour le LTE (liaison montante et descendante) est faite. Par conséquent, nous offrons notre évaluation et les critiques.**

## I. INTRODUCTION

Long Term Evolution (LTE), ou systèmes 3.9G, conçu à l'origine pour permettre d'atteindre des débits de données important (50Mbit/s dans le sens montant Uplink et 100Mbit/s dans le sens descendant Downlink dans une bande de 20 MHz), tout en permettant de minimiser la latence en offrant un déploiement flexible de la bande passante. Il est désigné comme le successeur des réseaux 3G. Il permet la bonne exécution des services internet émergents ces dernières années. Il utilise la commutation de paquet tout comme les réseaux 3G, à la différence qu'il utilise le multiplexage temporel (TD) et le multiplexage fréquentiel (FD) en même temps ce qui n'est pas le cas par exemple du HSPA qui n'effectue que le multiplexage temporel, ceci nous permet d'avoir un gain de débit (en efficacité spectral) d'environ 40%. [1]

LTE utilise L'OFDMA (Orthogonal Frequency Division Multiple Access), comme méthode d'accès dans le sens descendant (Downlink) (eNodeB→UE), elle combine TDMA et FDMA. Elle est dérivée du multiplexage OFDM, mais elle permet un accès multiple en partageant les ressources radio entre plusieurs utilisateurs. Son principe est de diviser la bande totale en multiples sous bandes orthogonales de taille étroite, ce processus permet de lutter contre le problème des canaux sélectifs en fréquences, ISI (Inter Symbol Interference), en plus, elle permet pour une même largeur spectrale, un débit binaire plus élevé grâce à sa grande efficacité spectrale (nombre de bits transmis par Hertz) en plus de sa capacité à conserver un débit élevé même dans des environnements défavorables avec échos et trajets multiples des ondes radio. Pour le sens ascendant (Uplink), la méthode utilisée est SC-FDMA, une variante de l'OFDMA, elles ont pratiquement les mêmes performances (débit, efficacité…etc.), mais SC-FDMA transmet les sous bandes séquentiellement pour minimiser le PAPR (Peak-to-Average Power Ratio, OFDMA a un grand PAPR), ceci est nécessaire, car pour le sens (UE→eNodeB), l'équipement terminal est doté d'une batterie d'une durée de vie limitée.

Un élément important dans l'architecture LTE, il se situe spécifiquement dans l'eNB, le RRM (Radio Resource Management), constitué principalement de deux tâches AC (**A**dmission **C**ontrol) et PS (Packet Scheduler).

L'AC est responsable de l'acceptation et du rejet des nouvelles requêtes, par contre le PS réalise l'allocation des ressources efficacement aux différents utilisateurs déjà acceptés par l'AC.

L'AC, traite les nouvelles demandes de connexion au réseau, la décision d'accepter ou de rejeter une requête dépend de la capacité du réseau à offrir la QoS exigée par cette requête tout en assurant la QoS des requêtes déjà admises dans le système.

Le PS quant à lui effectue le mapping UE-RB, c'est-à-dire. Sélectionner les utilisateurs UEs qui vont utiliser le canal en leurs affectant les ressources radios RBs qui leurs permettent de maximiser au plus les performances du système.

Il existe plusieurs paramètres pour évaluer les performances du système, par exemple on peut citer : l'efficacité spectrale (débit total du système), l'équité entre les UEs, le temps d'attente de chaque UE avant qu'il soit servis. La diversité des paramètres de performances a permis la création de plusieurs types d'ordonnanceurs.

Un paramètre important dans la conception des ordonnanceurs est la prise en charge de la QoS. Ceci a obligé le réseau LTE à faire la distinction entre les flux de données et donc on distingue :

*Classe conversationnelle* : c'est la classe la plus sensible aux retards et délais, elle comporte la vidéo conférence et la téléphonie. Elle ne tolère pas les délais car elle suppose que sur les deux extrémités de la connexion se trouve un humain.

*Classe streaming* : semblable à la classe précédente, mais elle suppose qu'une seule personne se trouve au bout de la connexion, de ce fait, elle est moins contraignante en terme de délais et de retards. Par exemple : vidéo-streaming

*Classe interactive* : des exemples de cette classe peuvent êtres : navigation web, accès aux bases de données …etc.

A l'inverse des types précédemment cités, les données doivent êtres délivrées dans un intervalle de temps, mais ce type de trafic met l'accent sur le taux de perte des données (Packet Error Rate).

*Classe Background* : appelé aussi classe des flux Best Effort, aucune QoS n'est appliquée, elle tolère les délais, la perte des paquets. Des exemples de cette classe : FTP, E-mails etc.… [2]

Deux autres paramètres influent sur la conception des algorithmes d'ordonnancement en LTE Uplink. Ces deux paramètres sont imposés par la méthode d'accès SC-FDMA, sont : la minimisation de la puissance de transmission (pour maximiser au maximum la durée de vie des batteries des UEs), en plus, les RBs alloués à un seul UE doivent être contiguës. Ceci rend l'allocation des ressources radio pour LTE Uplink plus difficile que celle pour le Downlink.

Le reste du papier est organisé comme suit : dans la section II, sera présentée la modélisation mathématique du problème d'allocation de ressources radio, dans la section III, un état de l'art sera présenté sur les algorithmes d'ordonnancement existant dans la littérature, on évaluera les performances de ces algorithmes avec quelques critiques dans la section VI, puis une conclusion et des perspectives seront présentés dans la section V.

II. CARACTERISTIQUES SYSTEMES

Dans cette section on commencera par donner l'architecture du LTE, puis on présentera la formulation mathématique du problème d'allocation de ressources radio.

A. Architecture LTE

L'architecture générale du LTE se comporte essentiellement du EPS (Evolved Packet System) qui comporte: le réseau cœur EPC (Evolved Packet Core) et la partie radio du réseau.

EPC consiste en un ensemble d'éléments de contrôle : MME (Mobility Management Entity), HSS (Home Subscriber Server), S-GW et P-GW (Serving et Packet-data Gateway). L'EPC est le responsable de la connexion avec les autres réseaux 3GPP et non 3GPP. La partie radio du réseau est composée des eNodeB (Enhanced NodeB) et UE (User Equipment). [3]

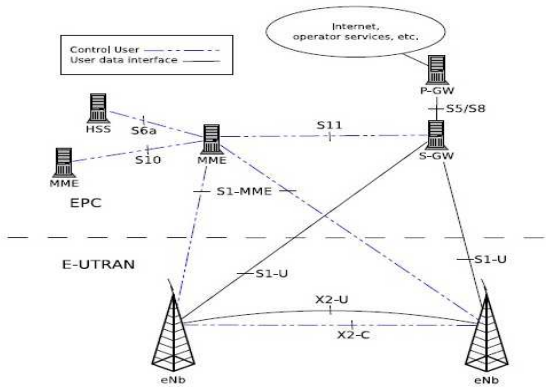

Figure-1 Architecture LTE [3]

B. Modélisation mathématique

On considère un système LTE où il y a N SBs (Scheduling Blocks c'est minimale de ressource allouée à un utilisateur est SB qui représente deux Resource Block RB consécutifs). avec une puissante égale partagée sur tous les SBs, en plus il y a K utilisateurs et le débit minimal demandé par le k-eme utilisateur est $R_k$ Mbit/s. on définie un SB comme un ensemble de $N_s$ symbole OFDM dans le domaine de temps TD et $N_{sc}$ sous porteuse dans le domaine de fréquence FD, en plus, en raison des signaux de contrôles et d'autres pilots, seulement $N_{sc}^d(s)$ des $N_{sc}$ sous porteuses seront utilisées pour transmettre les données du s-éme symbole OFDM, avec $s \in \{1,2,…,N_s\}$ et que $N_{sc}^d(s) \leq N_{sc}$. Supposant aussi $j \in \{1,2,…,J\}$ avec J est le nombre total de MCS (Modulation and Coding Scheme) supporté, alors soit $R_j^{(c)}$ le code associé au MCS j, $M_j$ est la constellation du MCS j et $T_s$ est la durée du symbole OFDM, alors le débit $r^{(j)}$ atteignable par un seul SB est :

$$r^{(j)} = \frac{R_j^{(c)} \log_2(M_j)}{T_s N_s} \sum_{s=1}^{N_s} N_{sc}^d(s) \qquad (1)$$

Maintenant, on définie $g_{k,n}$ comme CQI (Channel Quality Indicator) de l'utilisateur k sur le n-éme SB. Le CQI du k-éme utilisateur sur les N SBs est $g_k = [g_{k,1}, g_{k,2}, …, g_{k,N}]$ et pour tous les utilisateurs sur tous les BSs $G = [g_1, g_2, …, g_K]$. Le CQI est définie suivant le schéma de modulation, codage du canal.

Le $g_{k,n}$ est renvoyé par l'utilisateur k à la station de sase (eNb) pour que l'ordonnanceur détermine quel MCS doit être sélectionné pour le n-éme SB associé à l'utilisateur k.

Pour l'utilisateur k, La valeur maximale CQI sur tous les SBs est :

$$n^* = \arg\max(g_{k,n})_{n \in N} \qquad (2)$$

Par la suite, on définie $q_{k,\max(g_{k,n^*})} \in \{1,2,…,J\}$ la plus grande valeur du MCS atteinte par l'utilisateur k sur le n-éme SB pour la valeur CQI $g_{k,n^*}$, c'est-à-dire

$$q_{k,\max(g_{k,n^*})} = \arg\max(R_j^{(c)} \log_2(M_j) | g_{k,n^*}) \qquad (3)$$

Aussi il ne faut pas oublier le fait qu'un SB est alloué à un et un seul utilisateur, pour cela on définit $\rho_{k,n}$ l'indicateur d'allocation de ressource pour l'utilisateur k sur le n-éme SB, si $\rho_{k,n}=1$ alors le SB n est alloué à l'utilisateur k et que $\rho_{k',n} = 0$ pour tous $k' \neq k$

Posons $b_{k,j}$ le MCS choisi par l'utilisateur k sur tous les SBs qui lui sont alloués, $b_{k,j}=1$ veut dire que MCS j est choisi par l'utilisateur k.

Le débit atteint par l'utilisateur k sur une seule sous trame est :

$$r_k = \sum_{n=1}^{N} \rho_{k,n} \sum_{j=1}^{q_{k,\max(g_{k,n^*})}} b_{k,j} r^{(j)} \qquad (4)$$

Donc, le problème d'allocation de ressources radio a pour but de maximiser le débit des utilisateurs sous les contraintes suivantes :

$$\max_{\rho_{k,n}, b_{k,j}} \sum_{k=1}^{K} r_k \qquad (5)$$

Sous contrainte :

$$r_k \geq R_k \qquad \forall k \qquad (6)$$

$$\rho_{k,n} = 1, \rho_{k',n} = 0 \quad \forall k \neq k' \qquad (7)$$

$$\sum_{j=1}^{q_{k,max}(g_{k,n^*})} b_{k,j} = 1 \qquad (8)$$

(4) est la fonction objectif qui vise à maximiser le débit, (5) est une fonction visant a assuré un débit minimal pour tous les utilisateurs (c.-à-d. assuré une certaine QoS), (6) assure qu'un SB peut être attribué à un seul utilisateur, (7) tous les SBs d'un utilisateur emploient un seul MCS (contrainte sur les réseaux LTE). [1]

Le problème (4) est un problème NP-complet, par la suite plusieurs auteurs ont proposés leurs algorithmes qui visent à résoudre le (4).

### III. Ordonnancements en LTE

Dans cette section, on présentera un état de l'art sur les algorithmes d'ordonnancement existant pour les deux sens downlink et uplink. Ces algorithmes se basent sur les formulations mathématiques déjà mentionnées, essayent de réaliser l'allocation de ressources radio aux utilisateurs du système d'une façon efficace.

#### A. Algorithmes d'ordonnancement en Downlink

Les algorithmes d'allocation de ressources radio ont pour objectif d'améliorer les performances du système en augmentant l'efficacité spectral et l'équité dans le réseau. Il est donc essentiel de trouver un compromis entre l'efficacité (augmentation en débit) et l'équité entre les utilisateurs.

Plusieurs familles ou catégories d'algorithme existent dans la littérature, généralement chaque famille contient un ensemble d'algorithmes qui ont des caractéristiques communes.

*1) Les algorithmes opportunistes*

Ce type d'algorithme utilise des files d'attentes infinies, ces files d'attente sont utilisées dans le cas de trafic non temps réel. L'objectif principal de ce type d'algorithmes est de maximiser le débit global du système. Plusieurs algorithmes utilisent cette approche comme : PF (Proportional Fair), EXP-PF (Exponential Proportional Fair) etc...

*a) Proportional Fair (PF)*

Son but est d'essayer de maximiser le débit global du système en augmentant le débit de chaque utilisateur en même temps, il essaye de garantir l'équité entre les utilisateurs [10], la fonction objectif représentant l'algorithme PF est :

$$a = \frac{d_i(t)}{d_i^-} \qquad (9)$$

$d_i(t)$ : Débit correspondant au CQI de l'utilisateur i.
$d_i^-$ : Débit maximum supporté par le RB.

*b) Exponential Proportional Fair (EXP-PF)*

C'est une amélioration de l'algorithme PF qui supporte les flux temps réel (multimédia), au fait, il priorise les flux temps réel par rapport aux autres [11]. Un utilisateur $k$ est désigné pour l'ordonnancement suivant la relation suivante :

$$k = \max_i a_i \frac{d_i(t)}{d_i^-} \exp(\frac{a_i W_i(t) - X}{1+\sqrt{X}}) \qquad (10)$$

$$X = \frac{1}{N}\sum_i a_i W_i(t) \qquad (11)$$

$W_i(t)$ : Délai toléré par le flux.
$a_i$ : Paramètre strictement positive pour tous $i$.

*2) Les algorithmes équitables*

Plusieurs travaux de recherches ont visé l'équité entre les utilisateurs dans les réseaux LTE, ces algorithmes présentent généralement une insuffisance au débit. A noter que l'équité ne veut pas dire l'égalité,

*a) Round Robin*

C'est une stratégie classique d'allocation des ressources radio, l'algorithme alloue la même quantité de ressource aux utilisateurs en partageant le temps, par conséquent, le débit diminue considérablement vue, que tous les utilisateurs du système utilisent les ressources radio suivant un quantum de temps.

*b) Max-Min Fair (MMF)*

L'algorithme distribue les ressources entre les utilisateurs successivement en vue d'augmenter le débit de chaque utilisateur. Une fois que l'utilisateur alloue les ressources demandées pour atteindre son débit, on passe à l'utilisateur suivant. L'algorithme s'arrête par épuisement des ressources ou que les utilisateurs soient satisfaits.

*3) Algorithmes considérant les délais*

Ce type d'algorithme traite les délais d'arriver et de délivrance des paquets. Conçue principalement pour traiter les flux temps réel (multimédia et VoIP). Si un paquet dépasse ces valeurs de retard toléré, il sera supprimé de liste des flux à ordonnancer ce qui dégrade considérablement la QoS. M-LWDF (Maximum-Largest Weighted Delay First) est un exemple d'implantation de cette famille.

*a) M-LWDF*

Cet algorithme prend en charge des flux ayant des exigences de QoS différentes, il essaye de pondérer les retards des paquets en utilisant la connaissance de l'état de canal, a un instant $t$, l'algorithme choisi un utilisateur $k$ pour l'ordonnancement via la formule :[12]

$$k = \max_i a_i \frac{d_i(t)}{d_i^-} W_i(t) \qquad (12)$$

C'est pratiquement la même formule de l'algorithme EXP-PF, sauf que $a_i = -\log(p_i)T_i$, avec

$p_i$ : La probabilité que le délai ne soit pas respecté.
$T_i$ : Le délai que l'utilisateur $i$ peut tolérer.

Cet algorithme s'adresse principalement au flux temps réel qui exige le respect des délais, il donne de bons résultats dans ce contexte, par contre pour les flux non temps réel, ce n'est vraiment pas un bon choix vu que le délai n'est vraiment pas un paramètre important.

*4) Algorithmes optimisant le débit*

Ce type d'algorithme essaye de maximiser la fonction objective qui représente le débit, cette approche traite les flux temps réel et non temps réel, l'allocation de ressources dépend de la taille de la file de chaque utilisateur. Exemple d'algorithme de cette famille EXP Rule, Max-Weight etc.

*5) Les algorithmes multi-classe*

Cette approche considère les classes de flux où le traitement est différent pour chaque classe RT et NRT. Ce type d'algorithme privilégie les flux temps réel par rapport aux non temps réel, ce qui le rend les plus adéquats et plus efficace pour l'ordonnancement en LTE, par contre l'équité n'est vraiment pas considérer.

### B. Algorithmes d'ordonnancement en Uplink

Contrairement à l'ordonnancement côté downlink, l'ordonnancement côté uplink est bien plus compliqué pour plusieurs raisons, premièrement, c'est l'UE qui envoie les données et nous savons très bien que l'UE est doté d'une source d'énergie limitée, deuxièmement, c'est très difficile de prévoir le nombre de ressources radio nécessaires à un UE pour qu'il puisse échanger ces données avec la station de base. Suivant la fonction objectif prise en considération et suivant les classes de trafic qui passe par-dessus les canaux radio, nous avons trois grandes catégories d'ordonnanceurs : ceux traitons les flux best-effort, ceux qui prennent en considération la QoS et ceux optimisant la puissance d'émission. Dans cette partie nous allons essayer de faire le tour sur les principaux familles d'algorithmes d'allocation de ressources en LTE uplink.

#### 1) Paradigmes utilisés

Pour l'allocation des ressources radio en LTE uplink, le PS a besoin d'une matrice d'association entre UE-RB en entrée pour pouvoir donner en résultat les meilleures combinaisons qui améliorent les performances du système.

Pour la création de cette matrice, il existe dans la littérature deux grands paradigmes (Channel Dependent CD et Proportional Fairness PF)

Le premier CD, dans le processus de la création de la matrice, CSI (Channel State Information) ou bien l'état du canal est considéré, donc UE qui ont les plus grandes valeurs CSI auront la chance d'allouer plus de ressources, cette approche atteint les meilleures valeurs en débit, mais elle souffre de problème de famine.

Tandis que le PF, quant à lui, il prend le rapport CSI sur le débit pour chaque UE. Donc l'équité est proportionnelle sur la valeur CSI de la matrice. Cette approche atteint de bons débits en résolvant en même temps le problème de famine. [4]

#### 2) Modélisation du système LTE uplink

Les algorithmes d'ordonnancement en uplink prennent en entré une matrice avec K lignes (nombre d'UEs actifs) et M colonnes (nombre de RBs). $M_{i,m}$ est la valeur associée en UE i et RB m. Suivant le paradigme utilisé, cette valeur représente le CSI (Channel State Information) de chaque RB pour chaque UE, ou le rapport CSI sur débit.

Les valeurs de la matrice représentent l'association entre UE-RB, ces valeurs sont utilisées par l'ordonnanceur.[4]

#### 1) Ordonnaceurs de flux best effort

L'objectif principal de ce type d'ordonnanceur est, de maximiser l'utilisation des ressources radio dans le système et/ou l'équité du partage des ressources entre les UEs. Comme nous avons déjà dit, chaque algorithme à une fonction objectif à optimiser, ce type d'algorithme utilise une métrique PF.

Parmi les anciens travaux existant sur ce type d'ordonnanceurs (best-effort), on trouve les algorithmes gloutons, ils sont très efficaces pour ce genre de trafic (best effort).

|  | $RB_1$ | $RB_2$ | …. | $RB_M$ |
|---|---|---|---|---|
| $UE_1$ | $M_{1,1}$ | $M_{1,2}$ | … | $M_{1,M}$ |
| $UE_1$ | $M_{2,1}$ | $M_{2,2}$ | … | $M_{2,M}$ |
| . . . | . . . |  | … |  |
| $UE_K$ | $M_{K,1}$ | $M_{K,2}$ | … | $M_{K,M}$ |

Figure 2- Matrice d'association UE-RB

Cet algorithme utilise la métrique PF et il essaye de maximiser la fonction objective suivante :

$$U = \sum_{u \in U} \ln R(u) \quad (13)$$

R(u) : Débit moyen du UE u à l' instant t. L'utilisation de la fonction logarithme est pour avoir une équité proportionnelle.

Dans [5] les auteurs ont proposé trois algorithmes : FME (First Maximum Expansion), RME (Recursive Maximum Expansion) et MAD (Minimum Area Difference). Ces trois algorithmes appartiennent à la même catégorie (celle traitant les flux best-effort), c'est pour cette raison qu'ils ont la même fonction objective, mais ils se diffèrent de la manière dont les ressources sont allouées.

#### 2) Ordonnanceurs considérant la QoS

Deux paramètres importants dans la prise en considération de la QoS, sont le délai toléré et QoS de l'UE qu'on veut servir et les UEs déjà servies (suivant le type de flux).

Parmi les algorithmes proposés, il y a PFGBR (Proportional Fair with Guaranteed Bit Rate). Depuis son nom, on distingue deux métrique PF et GBR, la métrique PF est utilisée pour ordonnancer les UEs avec flux non GBR et pour ceux ayant un flux GBR, l'algorithme change la métrique pour pouvoir différencier les UE (donner des priorités aux UEs).[6]

$$M(u,c) \begin{cases} \exp\left(\alpha.\left(R_{GBR} - R^-(u)\right)\right).\dfrac{R^*(u,c)}{R(u)} & u \in U_{GBR} \\ \dfrac{R^*(u,c)}{R(u)} & u \in U_{non-GBR} \end{cases} \quad (14)$$

$R^-(u)$ : Débit moyen de l'utilisateur u au TTI t.

$R^*(u,c)$ : Débit estimé de l'utilisateur u, sur le Ressource Chunk c (RC ensemble continue de RB) au TTI t.

Les auteurs dans [7], ont proposés deux algorithmes qui considèrent la QoS. La fonction objective utilisé est défini comme suit :

$$\max \sum_{u \in U} \sum_{r \in RB} \alpha_{u,r}.\ f_r \quad (15)$$

$\alpha_{u,r}$ :=1 si le RB r est alloué au UE u.

$f_r$ est définit comme suit :

$$f_r = \frac{R_u * D_i^{max}}{R_i^{min} * D_i^{avg}} \qquad (16)$$

$R_u$ :   Débit atteignable.

$R_i^{min}$ : Débit minimal de la classe de service i.

$D_i^{max}$ : Délai max de la classe i.

$D_i^{avg}$ : Délai moyen de la classe i.

Le premier algorithme s'appelle SC-PS, Single Channel-Packet Scheduling, il réalise l'allocation d'un seul RB pour un UE donné dans un TTI. Dans le cas ou le nombre de UE u qui demande des ressources est inférieur au nombre de RB disponible, l'ordonnanceur distribue l'ensemble des RBs sur les UE équitablement $\frac{N_{RB}}{N_u}$. Dans le cas contraire, il fait l'allocation d'un RB au UE qui a les mauvaises conditions (par exemple : celui qui a le délai max est presque atteint) ainsi de suite. L'objectif principal de cet algorithme est d'allouer les ressources aux UEs ayant des contraintes de QoS plus sévère.

Le deuxième algorithme s'appelle MC-PS, Multiple Channel-Packet Scheduling, similaire au premier, a la différence que celui-là permet l'allocation de plusieurs RBs pour un seul UE. Cet algorithme a le même comportement dans le cas ou le nombre de UEs est inférieur au nombre de RBs disponible dans le système. Dans le cas où le nombre d'UE est supérieur à celui des RBs disponible, alors on fait l'allocation des $n = \left\lceil \frac{R_i^{min}}{R_u} \right\rceil$ RBs aux UEs suivant les valeurs de $f_r$ (on commence par ceux ayant les mauvaises conditions), on cherche tout d'abord le RB qui maximise le débit et puis on regarde à gauche et à droite de ce RB jusqu'à l'allocation des n RBs.

*3) Ordonnanceurs traitant la puissance du signal*

Le but principal de cette catégorie d'algorithmes est de minimiser la puissance du signal émis, pour essayer de rallonger la durée d'activité de l'UE, ce qui coïncide avec l'objectif de la méthode d'acces SC-FDMA . Cette approche n'a pas été vraiment trop traitée par les chercheurs, du coup il y a peu d'algorithmes dans la littérature. Citons par exemple les travaux [8]et [9].

IV.   DISCUSSION ET EVALUATION DES PERFORMANCES

- PF, est un ordonnanceur souvent utilisé dans les réseaux 3G, vu que le débit de ce type de réseaux est limité. Pour les réseaux après 3G, un facteur essentiel entre dans le jeu, c'est le délai surtout pour les flux multimédia qui représente le type de flux les plus importants dans les réseaux après 3G, ce facteur n'est pas pris en compte par cet algorithme, de ce fait, pour les flux non temps réel il fonctionne très bien par contre pour les flux temps réel il n'est pas préférable.
- Concernant le EXP-PF, les paramètres $W_i(t)$ et $a_i$ définissent le niveau de QoS requis par le flux. Ces paramètres essayent de donner plus d'importance aux applications ayant des un niveau de QoS plus élevé. Dans le cas ou la partie exponentielle de la formule soit égale a un, on retrouve la formule de l'algorithme PF. Ce cas de figure est possible si les flux ont pratiquement les mêmes délais pour les différents utilisateurs.
- En ce qui concerne le RR, il ne prend pas en considération la QoS, car les flux n'ont pas les mêmes besoins (VoIP, Streaming etc.), en plus alloué la même quantité de ressources n'est pas vraiment équitable, car les utilisateurs n'ont pas forcement les mêmes conditions de canal, ni les mêmes types de flux etc. Les réseaux après 3G, spécifiquement LTE se focalise sur la QoS des flux temps réel, du cout, utilisé RR n'est vraiment pas le bon choix.
- Le fait d'essayer de satisfaire tout les utilisateurs dans l'algorithme MMF, donne l'avantage aux utilisateurs ayant des faibles exigences qu'ils seront souvent servis, par contre on pénalise les utilisateurs qui demandent plus de ressources. Cette approche ne prenne pas en compte la diversité multiutilisateur et que les flux ont des exigences en QoS différentes et l'équité ne veut pas dire l'égalité. En résumé, cet algorithme n'est vraiment pas le bon choix pour l'ordonnancement en LTE.

Pour résumé, l'allocation de ressources radio est faisable (plusieurs algorithmes et approches existent), mais la diversité des flux (QoS) et les conditions radio affectent les performances de l'algorithme. L'allocation de ressource est un problème NP-complet, vu que l'algorithme essaye de maximiser et/ou minimiser plusieurs paramètres en même temps. Pour cette raison, chaque approche ou algorithme essaye d'optimiser le maximum des paramètres qu'il pourra.

Concernant le côté uplink, il est beaucoup plus compliqué vue les nouvelles contraintes imposées, comme, les RBs alloués à un seul utilisateur doivent êtres continues, plus la contrainte sur la puissance du signal émis. Les algorithmes traitant la QoS sont les plus adaptés et les plus sondés, car ils traitent le facteur le plus important dans les réseaux LTE, qui est la QoS des flux.

V.   CONCLUSION

L'allocation de ressource radio se fait dans l'eNB par le PS, cette tâche est trop complexe, car elle nécessite de prendre en considération plusieurs facteurs en même temps, en plus elle doit être immédiate (en temps réel).

L'objectif de cet article est de présenter un état de l'art sur l'allocation de ressources radio en LTE. Dans ce travail, nous avons essayé de faire le tour des approches existantes dans la littérature dans les deux sens downlink et uplink, nous avons aussi cité quelques algorithmes, nous avons montré les avantages et les inconvénients de chaque catégorie, par la suite il serait plus judicieux de se focaliser sur un type de trafic, essayer d'améliorer les performances, ça sera sans doute les flux temps réel.

REFERENCES

[1]   Raymond Kwan, C.leung et J.Zhang, "Downlink Resource Scheduling in an LTE System", IEEE Signal Processing letters, vol 16, no 16 pp 461-464, Jun 2009.

[2]   Khalid Elgazzar, Mohamed Salah, Abd-Elhamid M.Taha, Hossam Hossanein, "Comparing Uplink Schedulers for LTE".

[3] Geovany Mauricio Itturalde Ruiz, "Performances des réseaux LTE",Thése, Institut National Polytechnique de Toulouse (INP Toulouse),2012.

[4]   Haider Safa and Kamal Tohme,"LTE Uplink Scheduling Algorithms: Performance and Challenges", 19 International Conference on Telecommunications (ICT), 2012.

[5] L. Ruiz de Temino, G. Berardinelli, S. Frattasi, and P. Mogensen, "Channel-Aware Scheduling Algorithms for SC-FDMA in LTE Uplink," in Personal, Indoor and Mobile Radio Communications, 2008. PIMRC 2008. IEEE 19th International Symposium.

[6] P. Wen, M. You, S. Wu, and J. Liu, "Scheduling for streaming application over wideband cellular network in mixed service scenarios," in Military Communications Conference, 2007. MILCOM 2007. IEEE, pp. 1-7, Oct. 2007.

[7] Oscar Delgado, Bridgite jaumard,"Scheduling and Resource Allocation for Multiclass Services in LTE Uplink systems", 6th international conference on

wireless and mobile computing, networking and communications,2010.

[8] Z. Li, C. Yin, and G. Yue, "Delay-bounded power-e_cient packet scheduling for uplink systems of lte," in Wireless Communications, Networking and Mobile Computing, 2009. WiCom '09. 5th International Conference on, pp. 1 {4, Sept. 2009.

[9] F. Sokmen and T. Girici, "Uplink Resource Allocation Algorithms for Single-Carrier FDMA Systems," in Wireless Conference (EW), 2010

European, pp. 339-345, Apr. 2010.

[10] R. Kwan; C. Leung; and J. Zhang. Proportional fair multiuser scheduling in lte. In IEEE Signal Processing Letters, Vol. 16(6), pages 461–464, June 2009.

[11] R. Basukala; H. Mohd Ramli and K. Sandrasegaran. "Performance analysis of exp/pf and m-lwdf in downlink 3gpp lte system". In IEEE First Asian Himalayas Conference, Vol. 1, pages 1–5, November 2009.

[12] P. Ameigeiras; J. Wigard and P. Mogensen. Performance of the mlwdf scheduling algorithm for streaming services in hsdpa. In IEEE Vehicular Technology Conference (VTC), Vol. 2, pages 999–1003, September 2004